# The Number of Information Bits Related to the Minimum Quantum and Gravitational Masses in a Vacuum Dominated Universe

Ioannis Haranas[1] Ioannis Gkigkitzis[2]

[1] Department of Physics and Astronomy, York University
4700 Keele Street, Toronto, Ontario, M3J 1P3, Canada
E-mail: yiannis.haranas@gmail.com

[1] Departments of Mathematics, East Carolina University
124 Austin Building, East Fifth Street, Greenville
NC 27858-4353, USA
E-mail: gkigkitzisi@ecu.edu

**Abstract**
Wesson obtained a limit on quantum and gravitational mass in the universe by combining the cosmological constant $\Lambda$, Planck's constant $\hbar$, the speed of light $c$, and also the gravitational constant $G$. The corresponding masses are $2.0 \times 10^{-62}$ kg and $2.3 \times 10^{54}$ kg respectively, and in general can be obtained with the help of a generic dimensional analysis, or from an analysis where the cosmological constant appears in a four dimensional space-time and as a result of a higher dimensional reduction. In this paper our goal is to establish a relation for both quantum and gravitational mass as function of the information number bit $N$. For this reason, we first derive an expression for the cosmological constant as a function of information bit, since both masses depend on it, and then various resulting relations are explored, in relation to information number of bits $N$. Fractional information bits imply no information extraction is possible. We see, that the order of magnitude of the various parameters as well as their ratios involve the large number $10^{122}$, that is produced naturally from the fundamental parameters of modern cosmology. Finally, we propose that in a complete quantum gravity theory the idea of information the might have to be included, with the quantum bits of information ($q$-bits) as one of its fundamental parameters, resulting thus to a more complete understanding of the universe, its laws, and its evolution.

**Key words**: Cosmological constant, quantum mass, gravitational mass, information bit, fractional information bit, large number hypothesis.

## 1. Introduction

Observational data from galaxies and gravitational lensing of high $z$ quasars and cosmic microwave background radiation, suggest that the 99% of the universe material consists of dark matter (Overduin, and Wesson, 2003). Therefore the density of the vacuum contributes a high fraction of the corresponding dark matter, whose energy density is given by:

$$\rho_v = \rho_\Lambda = \frac{\Lambda c^2}{8\pi G} \tag{1}$$

___
In this paper all authors have contributed equally.

where $\Lambda$ is the cosmological constant, $c$ is the speed of light and $G$ is the gravitational constant. Today's data indicate that $\rho_\Lambda = 6.0 \times 10^{-30}$ g/cm$^{-3}$ and therefore $\Lambda = 1.117 \times 10^{-52}$ m$^{-2}$ (Krauss and Starkman, 1999), which makes that the distance to the horizon approximately is $R_H \cong 1.70 \times 10^{26}$ m (Krauss and Starkman, 1999). It is now common knowledge that the cosmological constant of general relativity is a parameter derived out of higher-dimensional theories in a four dimensional reduction, in an effort to unify gravity with particle physics (Weston, 1999). Some of these theories include 10$D$ supersymmetry, 11$D$ supergravity and 26$D$ string theories. The aforementioned theories should present the natural ground where the quantum of action $\hbar$ should naturally appear. In particular, theories like the basic extension of 4$D$ Einstein theory and the low-energy limit of higher-$D$ theories constitute the modern incarnation of (non-compact) 5$D$ Kaluza-Klein theory. This has been intensively studied recently, under the names induced-matter theory (Wesson, 1992) and (Mashhoon, et al., 1998), and membrane theory (Randall, Sundrum, 1998) and (Arkani-Hamed, 1998). Both theories predict the existence of a fifth force, which might be the force via with particles can interact (Youm, 2000). Therefore and as a result it might be possible to detect massive, and massless or photon like particles as well as photons particles in the spacetime.

Following Wesson (2004) two different mass scales can be formed. First, a quantum mass scale given by:

$$m_{qu} = \frac{\hbar}{c}\sqrt{\frac{\Lambda}{3}} = 2.0 \times 10^{-62} \text{ kg}, \qquad (2)$$

where and $\Lambda$ is the cosmological constant, $\hbar$ is Planck's constant, and c is the speed of light. This is the scale for the minimum quantum mass in the universe. Similarly, the gravitational mass scale is given according to the equation (Wesson, 2004):

$$m_{gr} = \frac{c^2}{G}\sqrt{\frac{3}{\Lambda}} = 2.3 \times 10^{54} \text{ kg}, \qquad (3)$$

where $G$ is the constant of universal gravitation. Our goal is to introduce the idea of information as a new potential parameter in many of today's natural phenomena. In particular, this contribution, investigates Eqs (2) and (3) given in Wesson's original paper and their relation to the number of information bits, via a relation that relates the cosmological constant $\Lambda$ to the number of information bits $N$. Next, we further investigate various resulting expressions and their relation to the number of information bits $N$. As an example, Planck's length $\ell_P$ is expressed in terms of the quantum particle's Compton wavelength and the number of information bits $N$. Next, the number of information bits $N$ involved in both particle masses as defined by Wesson is derived. Finally, using the *Margolus-Levitin* theorem, we calculate the number of operations performed by the two particles at time *t* equal to the age of the universe as a function of the number of information bits.

## 2. Quantum and Gravitational Masses and the Number of Information Bits

It is now a suggestion that quantum mechanics is non-local, but at the same time a fundamental mechanism for that is not known. However, the holographic principle indicates a possible non-locality mechanism in any vacuum-dominated Friedmann universe. To be more precise, a holographic non-local quantum mechanical description can be possible for a finite amount of information in a closed vacuum-dominated universe. Today's theories, assume that the universe began by a quantum fluctuation from nothing, underwent inflation and became so large that it is locally almost flat, and that after the inflationary era the vacuum energy density of the universe is constant. This is the case of the existence a non-zero cosmological constant $\Lambda$. More information on such a universe arising a in quantum cosmological way is presented in Mongan (2001). When such closed universe began it already contained all the information that it will ever contain. If nothing exists outside of the closed physical universe, that would imply that no information can come into the universe from elsewhere either (Mongan 2001). In the case of a de-Sitter metric we solve for the gravitational radius of such universe we have that:

$$\left(1 - \frac{\Lambda}{3} r^2\right) = 0, \tag{4}$$

and the de-Sitter horizon can be easily found to be:

$$r_{H_{1,2}} = \pm \sqrt{\frac{3}{\Lambda}}. \tag{5}$$

One of the most enigmatic features of de Sitter space is its entropy. That de Sitter space has finite entropy may be expected, based on the appearance of a horizon in pure de Sitter. This horizon is, however, qualitatively different from a black hole horizon. The position of the horizon is observer dependent, and because of this it is not entirely clear which concepts about black holes carry over to de Sitter space. In fact, the de Sitter cosmological horizon looks in many ways like the Rindler horizon in Minkowksi space. Let us now proceed with the derivation of how the cosmological constant $\Lambda$ depends on the number of information bits *N*. We start with the entropy of such black hole that can be written as according Bekenstein (1973) to be:

$$S = \frac{k_M}{4\ell_p^2} A_H \tag{6}$$

This is the Bekenstein-Hawking area-entropy law, which says that the entropy *S* is associated with an event horizon, and where $k_B$ is the Boltzmann constant, is the horizon area $A_H$ divided by $4\ell_p^2$, where

$\ell_p^2 = \frac{G\hbar}{c^3}$ is Planck's length (Bekenstein, 1973). This is a macroscopic formula and it should be viewed in the same light as the classical macroscopic thermodynamic formulae. It describes how properties of event horizons in general relativity change as their parameters are varied. Therefore the entropy of such a black is given by (Bousso and DeWolfe, 2002):

$$S_{U_H} = \frac{k_B}{4\ell_P^2} A_H = \frac{k_B}{4\ell_P^2} \left( \frac{2\pi^{\frac{n}{2}}}{\Gamma\left(\frac{n}{2}\right)} \left( \frac{(D-1)(D-2)}{2\Lambda} \right)^{\frac{(D-2)}{2}} \right) = \frac{k_B}{4\ell_P^2} \left( \frac{12\pi}{\Lambda} \right) = \frac{3\pi k_B}{\Lambda \ell_P^2} \quad (7)$$

where $D = 4$, is the dimensionality of the de-Sitter space, $n$ represents points on $n-1$ dimensional sphere and $\Gamma$ is the gamma function of the indicated argument. Solving for the cosmological constant $\Lambda$ we obtain:

$$\Lambda = 3\pi \left( \frac{k_M}{S_{U_H} \ell_P^2} \right), \quad (8)$$

Taking into account that the number of information bits $N$ relates to the entropy $S$ in the following way is given (Lioyd, 2000):

$$S = k_B N \ln 2 \quad , \quad (9)$$

where $k_B$ is the $S$ is the Boltzmann constant. Substituting in Eq. (8) we obtain that:

$$\Lambda = \left( \frac{3\pi}{N \ln 2 \ell_P^2} \right) = \frac{3\pi}{N \ln 2} \Lambda_{max} . \quad (10)$$

where $\Lambda_{max} = 1/\ell_P^2 = c^3/G\hbar$ (Haranas, 2002) the maximum cosmological constant, $N$ the number of information bits, $\ell_p$ is the Planck length. This is the cosmological constant as a function of the information bits associated with the cosmological horizon. Our derived expression agrees with the equation given in Mongan (2007). Next substituting Eq. (10) in Eq. (2) and rearranging we obtain that and equations which give the dependence of Wesson masses on the number of information bits $N$, we obtain the following expressions:

$$m_{qu} = \frac{\hbar}{\ell_p c} \sqrt{\frac{\pi}{N \ln 2}} = 4.530 \left( \frac{\hbar}{\ell_p c \sqrt{N}} \right), \quad (11)$$

solving for the Planck length we can express this important constant of physics as a function of the information bit in the following way:

$$\ell_P = \frac{\hbar}{m_{qu} c} \sqrt{\frac{\pi}{N \ln 2}} = 4.530 \left( \frac{\lambda_{qu_{Com}}}{\sqrt{N}} \right), \quad (12)$$

where $\lambda_{qu_{Com}} = \hbar/m_{qu}c$ is the Compton wavelength related to the minimum quantum mass scale particle in the universe as it is given in Wesson (2004). Next, solving for the information bit number $N$ related to the minimum quantum mass we obtain that:

$$N_{qu} = 20.521 \left( \frac{\lambda_{qu_{Com}}}{\ell_P} \right)^2 = 20.521 \lambda_{qu_{Com}}^2 \Lambda_{max}, \qquad (13)$$

therefore we find that the number of information bits involved in Wesson's minimum quantum mass is given by the square of the ratio of its corresponding Compton wavelength over the Planck length. Similarly, for the gravitational mass in the universe we obtain the following information bit number $N$ relation:

$$m_{gr} = \frac{c^2}{G} \sqrt{\frac{3N\ell_P^2 \ln 2}{3\pi}} = 0.469 \left( \frac{c^2 \ell_P}{G} \right) \sqrt{N}, \qquad (14)$$

solving for the number of information bits $N$ we obtain that:

$$N_{gr} = 4.546 \left( \frac{Gm_{gr}}{c^2 \ell_P} \right)^2 = 4.546 \left( \frac{R_{m_{gr}}}{2\ell_P} \right)^2 = 4.546 \left( \frac{\Lambda_{max}}{\Lambda} \right). \qquad (15)$$

where $R_{m_{gr}}$ is the gravitational radius of the corresponding gravitational mass. Equation (9) demonstrates that the number of information bits $N$ involved in the gravitational mass $m_{gr}$ is equal to the ratio of half the particle's gravitational radius over its wavelength Compton wavelength. From Eqs. (7) and (9) we have that:

$$\left( \frac{N_{qu}}{N_{gr}} \right) = 18.056 \left( \frac{R_{m_{gr}}}{\lambda_{qu_{Com}}} \right)^2. \qquad (16)$$

Let us now assume that the quantum mass $m_{qu}$ can be equal to the Planck mass. This will presuppose the following condition on the cosmological constant $\Lambda$. To find the condition on lambda we equate the Planck mass to the quantum mass derived by Wesson (2004) and therefore we have:

$$\sqrt{\frac{\hbar c}{G}} = \frac{\hbar}{c} \sqrt{\frac{\Lambda}{3}}, \qquad (17)$$

from which we obtain that:

$$\Lambda = \frac{3c^3}{G\hbar} = 3\Lambda_{max}, \qquad (18)$$

where $\Lambda_{max}$ is the value of the cosmological constant during the Planck era (Haranas, 2002). A cosmological constant $\Lambda=3\Lambda_{max}$ presupposes a time era slightly earlier that the Planck time limit, an era the equations of physics fail to describe. Similarly, requesting that $m_{gr} = m_{Pl}$ results in the same equation like Eq. (18), which again corresponds to an era earlier that the Planck era.

According to the *Margolus-Levitin* theorem, the maximum rate at which logical operation could be performed by a physical system with energy $E$ is $2E/\pi\hbar$ (Margolus and Levitin, 1998). Therefore, the maximum number of operation that could have been performed by the masses of the particles predicted by Wesson in the observable universe when written as a function of the number of information bits $N$ become. For the quantum mass $m_{qu}$ limit we obtain:

$$n_{m_{qu}} = \frac{2m_{qu}c^2}{\pi\hbar} t_u = \frac{2c}{\pi H_0}\sqrt{\frac{\Lambda}{3}} = \frac{2}{\pi}, \tag{19}$$

where $t_u = 1/H_0$ is the age of the universe, and $H_0 = c\sqrt{\frac{\Lambda}{3}}$ (Islam, 1992) is Hubble's constant. We see that Eq. (19) proves to be independent of the number of information bits $N$. Similarly, for the gravitational mass $m_{gr}$ we obtain the following total number of operations that the gravitational mass can perform in a time equal to the age of the universe to be:

$$n_{m_{gr}} = \frac{2c^2}{\pi\hbar}\frac{c^2}{G}\sqrt{\frac{3}{\Lambda}} t_u = \frac{2c^4}{\pi\hbar G}\frac{1}{H_0}\sqrt{\frac{3}{\Lambda}} = \frac{6c^3}{\pi G\hbar\Lambda} = \frac{6}{\pi}\left(\frac{\Lambda_{max}}{\Lambda}\right), \tag{20}$$

which, in terms of the number of information bits $N$, the number of operations $n_{m_{gr}}$ becomes:

$$n_{m_{gr}} = \frac{6c^3}{\pi G\hbar}\left(\frac{N\ell_p^2 \ln 2}{3\pi}\right) = \frac{2N}{\pi}\ln 2. \tag{21}$$

Next, following Faus (2010) and using an expression that connects the cosmological constant to the Hubble parameter we can derive an expression of how the Hubble parameter relates to the information bit number $N$. Equating Eq. (10) to $\Lambda = 21\left(\frac{H_0}{c}\right)^2 = \frac{21}{c^2 t_0^2}$ (Faus, 2010) and (Islam, 1992) we obtain that

$$H_0 = \pm 0.805 \frac{c}{\ell_p\sqrt{N}} = \pm\frac{0.805}{t_p\sqrt{N}} \tag{22}$$

We obtain an $N^{1/2}$ dependence of the present value of the cosmological constant on the information bit number $N$. Using the time dependence of the cosmological constant as it is given by Faus (2010) we derive the dependence of information number bit $N$ on cosmic time $t$, and therefore we obtain the following expression:

$$N \cong \frac{3\pi}{\ln 2}\left(\frac{t}{t_P}\right)^2 = 4.681\times 10^{87} t^2 \tag{23}$$

where $t_p = (\hbar G/c^5) = 5.39\times10^{-44}$ s is the Planck time. Similarly, with reference to Hajduković (2010) the mass of the pion $m_\pi \cong (\hbar^2 H/Gc)^{1/3}$ (Hajduković, 2010) we obtain that the mass of the pion $m_\pi$ depends on the number of information bit and time in the following way:

$$m_\pi^3 \cong 0.805\left(\frac{\hbar^2}{G\ell_P\sqrt{N}}\right) = 0.218\left(\frac{\hbar^2 t_p}{G\ell_p}\right)t^{-1} \cong 0.218\left(\frac{\hbar^2}{Gc^2}\right)t^{-1}, \tag{24}$$

where $t_P = (\hbar G/c^5)^{1/2}$ is the Planck time, and $G$ is the gravitational constant. Therefore we find that the following pion mass dependences on the information bit $N$ and time $t$ are $m_\pi \propto N^{-3/2} \propto t^{-1/3}$. Next, with reference to Santos (2010), and using his derived expression for the predicted by him gravitational density given by the equation Santos (2010):

$$\rho_{grav} = \alpha_0 \frac{Gm^6}{\hbar^4}, \tag{25}$$

where $\alpha_0 = -4/3\pi^3$ can be written as function of information bit $N$ in the following way

$$\rho_{grav} = \frac{0.648\alpha_0}{G\ell_p^2 N} = 0.648\alpha_0 \frac{\Lambda_{max}}{GN} = \frac{0.047\alpha_0}{Gc^4 t^2} \tag{26}$$

And therefore we find that the gravitational density as derived by Santos (2010) depend on the information bit $N$ and time $t$ are $\rho_{grav} \propto N^{-1} \propto t^{-2}$.

## 3. Discussion and Numerical Results

To evaluate our findings let as use $m_{qu} = 2.3\times10^{-62}$ (Wesson, 2004) kg and from that we obtain the following Compton wavelength to be $\lambda_{qu_{Com}} = 1.756\times10^{19}$ m, or equivalently $\ell_{gr_{Com}} = 1.90\times10^{-8} R_{uni}$, and $\lambda_{gr_{Com}} = 1.527\times10^{-97}$ m therefore we obtain that the ratio of the quantum to gravitational Compton wavelengths is:

$$\frac{\lambda_{qu_{Com}}}{\lambda_{gr_{Com}}} = 1.150\times10^{116}. \tag{27}$$

Similarly, the corresponding number of information bits related to the quantum and gravitational mass scales are given by:

$$N_{m_{qu}} = 2.228\times10^{107} \text{ bits,} \tag{28}$$

$$N_{m_{gr}} = 1.563\times10^{22} \text{ bits}. \tag{29}$$

The corresponding relation between the numbers of bits numerically becomes:

$$\frac{N_{m_{gr}}}{N_{m_{qu}}} = 0.690 \times 10^{15}. \tag{30}$$

Furthermore, numerical values for the values of information bits involved in various resulting scenarios are calculated. We first calculate the amount of information bits involved in today's value of the cosmological constant. Using, the density of the vacuum to be $\rho_\Lambda = 6.0 \times 10^{-30}$ g/cm$^{-3}$ (Krauss and Starkman, 1999) and the distance to the horizon to be approximately $R_H \cong 1.70 \times 10^{26}$ m (Krauss and Starkman, 1999) we obtain, that the cosmological constant in the present era is $\Lambda = 1.117 \times 10^{-52}$ m$^{-2}$. Therefore substituting for $\Lambda$ we obtain that the number of information bits involved in the present value of the cosmological constant $\Lambda$ is:

$$N_\Lambda = \left(\frac{3\pi}{\Lambda \ln 2 \ell_p^2}\right) = 4.661 \times 10^{122} \text{ bits.} \tag{31}$$

Our predicted number agrees with that given in Funkhouser (2008). Similarly in the case where $\Lambda = \Lambda_{Max} = \ell_p^{-2} = \frac{c^3}{G\hbar} = 3.840 \times 10^{69}$ m$^{-2}$ (Haranas, 2002) we obtain that:

$$N_\Lambda = \left(\frac{3\pi}{\Lambda_{max} \ln 2 \ell_p^2}\right) = 13.40 \text{ bits,} \tag{32}$$

which corresponds to an entropy of approximately $S \approx 9.28 \, k_B$. Eq. (32) indicates a fractional number of information bits $N$. Since information bits can only take values of zero and one fractional information bits would imply no information extraction. Therefore taking the ratio of $N_\Lambda / N_{\Lambda_{max}}$ we find that:

$$\frac{N_\Lambda}{N_{\Lambda_{max}}} = 0.348 \times 10^{122}. \tag{33}$$

Finally, to find the number of information bits that the de-Sitter horizon contains we write Eq. (5) in terms of the number of information bits $N$ via the cosmological constant $\Lambda$ in the following way:

$$N = \frac{\pi}{\ln 2}\left(\frac{r_H}{\ell_p}\right)^2 = \frac{\pi}{\ln 2} \Lambda_{max} r_H^2 = 4.532 \times 10^{122} \text{ bits.} \tag{34}$$

First, in the case where the quantum mass $m_{qu}$ predicted by Wesson (2004) becomes equal to the Planck mass $m_{Pl}$, the cosmological constant $\Lambda$, is independent of the Planck length and involves a specific number of information bits that is just a numerical constant. Furthermore, in the case where the quantum mass $m_{qu}$ predicted by the above theories is equal to the Planck mass $m_{Pl}$ the cosmological constant $\Lambda$ involves a specific constant number of information bits that is independent of the Planck length and it is equal to:

$$N = \frac{\pi}{\ell_P^2 \ln 2}\left(\frac{G\hbar}{c^3}\right) = \frac{\pi}{\ln 2} = 4.532 \text{ bits}. \tag{35}$$

Here again fractional information bits correspond to now information. Similarly, for the ratio of gravitational to the quantum mass as predicted by Wesson (2004) we obtain that:

$$\frac{m_{gr}}{m_{qu}} = \frac{3c^3}{G\hbar\Lambda} = 3\left(\frac{\Lambda_{max}}{\Lambda}\right) = 0.429 \times 10^{122}. \tag{36}$$

In terms of the number of information bits $N$, Eq. (26) can be written as:

$$\frac{m_{gr}}{m_{qu}} = \frac{3}{\ell_{Pl}^2 \Lambda} = 3\left(\frac{\Lambda_{max}}{\Lambda}\right) = \frac{N \ln 2}{\pi} = 0.221N, \tag{37}$$

and therefore, the gravitational mass scale predicted by Wesson relates to the quantum mass scale in the following ways:

$$m_{gr} = 3\left(\frac{\Lambda_{max}}{\Lambda}\right) m_{qu}, \tag{38}$$

or equivalently in terms of the information bits $N$ the two mass scales relate in the following way:

$$m_{gr} = 3\left(\frac{N \ln 2}{\pi}\right) m_{qu}. \tag{39}$$

Similarly, the ratio of the mass of the universe to that of quantum mass as predicted by Wesson (2004) becomes:

$$\frac{M_{uni}}{m_{qu}} = 6.20 \times 10^{121}. \tag{40}$$

Finally, the numerical value of the total number of operations performed in the age of the universe by a gravitational particle is:

$$n_{m_{gr}} = 0.574 \times 10^{122}, \tag{41}$$

Therefore the their ratio is

$$\frac{n_{m_{gr}}}{n_{m_{qu}}} = 0.365 \times 10^{122}. \tag{42}$$

The number $10^{122}$ appears in an ensemble of pure numbers naturally produced from fundamental cosmological parameters that might constitute a new-large number coincidence similar to that postulated by Dirac. These numbers constitute a compelling, new large number coincidence problem (Funkhouser, 2008). In this paper, we demonstrate that new large number coincidence also exist in the relations involving the information bits $N$, and number of operations $n$, in its relation to the fundamental mass scales predicted by Wesson (2004), and also their relation to the cosmological constant $\Lambda$. All these are possible after a relation relating the cosmological constant $\Lambda$ to the information number bit $N$ is derived.

The involvement of the cosmological constant in many of today's cosmological relations introduces a direct relation of the cosmological parameters to the information number bit $N$. Certain cosmological scenarios involve fractional number of information bits, which implies no information extraction is possible. Our result is in agreement with Faus (2011). Few authors have given various explanations, for example, Funkhouser (2008) has demonstrated implicit physical pure number relations that result from the standard cosmological model. Our main interest is to express the basic relations in this paper as functions of the number of information bits $N$. Thus we have found that the minimum quantum mass and its corresponding Compton wavelength scale as $N^{-1/2}$, where the number of information bits involved in the minimum quantum mass scales as the ratio of the quantum Compton wavelength over the Planck length squared. Similarly, the gravitational mass scales as $N^{1/2}$, where the corresponding number of information bits is equal to the half of the gravitational radius of the gravitational mass divided by Compton wave length of the gravitational particle. It might be important that the number of information bits $N$ enters Wesson's definition of two different mass scales through the dependence of the cosmological constant $\Lambda$ that itself depends on the information bit $N$. Therefore the "gravitational information bit" and also the quantum of information bit also called the $q$-bit that is itself a microscopic system, such as an atom, or nuclear spin, or photon (Capurro and Hjørland, 2003), might be the corresponding units at these extreme scales of operation of nature's phenomena, where quantum gravity operates. It can be of a great interest to examine the quantum spacetime fluctuations from an information point of view, and try to place information boundaries on the ultraviolet cut off, but this will be our next paper. Thus an ultimate information theory might find its place at the heart of this quantum gravity theory, as well as in all the other theories mentioned above. In a universe that expands the number of information bits required to define a particle increase, and therefore the total amount of information also increases. If the cosmological constant varies as $\Lambda \cong 1/t^2$, this would result in an increased number of information bits, as the entropy of the universe increases. Therefore, at much later times in our universe a very large amount of information bounded by the light cones would be necessary to describe its evolution.

## 4. Conclusions

We have used the results predicted by Wesson in order to investigate the dependence of the minimum quantum and gravitational masses in a vacuum dominated Friedmann universe on the number of information bits $N$. The number $10^{122}$ appears in an ensemble of pure numbers naturally produced from fundamental cosmological parameters. Using our derived expression of the cosmological constant as a function of the information bit, we have found that the minimum quantum mass involves an $N^{-1/2}$ dependence on information bits, where the gravitational mass scale has an $N^{1/2}$ dependence respectively.

Finally, we propose that a complete quantum gravity theory might have to include the quantum bit of information as one of its fundamental parameters for a more complete description of the universe, its laws, and its evolution.

**Acknowledgements**: The authors would like to thank the editor in chief of Astrophysics and Space Science journal Dr. Michael Dopita, for encouraging us to resubmit a second revised version of our paper.